\documentclass[useAMS,usenatbib,usegraphicx]{mn2e}
\title[Are many radio-selected BL Lacs radio quasars in disguise?]{Are many radio-selected BL Lacs radio quasars in disguise?}

\author[V. D'Elia et al.]{V. D'Elia$^{1,2}$\thanks{E-mail:
delia@asdc.asi.it},  P. Padovani$^{3,4}$, P. Giommi$^{1}$, S. Turriziani$^{5}$ 
\\
$^1$ INAF-Osservatorio Astronomico di Roma, Via Frascati 33, I-00040 Monteporzio Catone, Italy;\\ 
$^2$ ASI-Science Data Centre, Via del Politecnico snc, I-00133 Rome, Italy;\\ 
$^3$ European Southern Observatory, Karl-Schwarzschild-Strasse 2, D-85748 Garching, Germany\\
$^{4}$ Associated to INAF - Osservatorio Astronomico di Roma, via Frascati 33,
I-00040
Monteporzio Catone, Italy\\
$^5$ University of Rome Tor Vergata, via della Ricerca Scientifica 1, I-00133, Roma, Italy
}
\begin{document}

\date{Accepted... Received...; in original form...}

\pagerange{\pageref{firstpage}--\pageref{lastpage}} \pubyear{2015}

\maketitle

\label{firstpage}

\begin{abstract}

  We show that a blazar classification in BL Lacs and Flat Spectrum
  Radio Quasars may not be adequate when it relies solely on the
  equivalent widths (EWs) of optical lines. In fact, depending on
  redshift, some strong emission lines can fall in the infrared window
  and be missed. We selected a sample of BL Lacs with firm redshift
  identification and good visibility from Paranal. We targeted with
  the X-shooter spectrograph the five BL Lacs with $z>0.7$, i.e.,
  those for which the H$\alpha$ line, one of the strongest among
  blazars, falls outside the optical window and determined the EW of
  emission lines in both the infrared and optical bands. Two out of
  five sources show an observed H$\alpha$ EW $>5$\AA\, (one has rest
  frame EW $> 5$\AA) and could be classified as FSRQs by one of the
  classification schemes used in the literature. A third object is
  border-line with an observed EW of $4.4\pm0.5$\AA. In all these
  cases H$\alpha$ is the strongest emission line detected. The
  H$\alpha$ line of the other two blazars is not detected, but in one
  case it falls in a region strongly contaminated by sky lines and in
  the other one the spectrum is featureless. We
  conclude that a blazar classification based on EW width only can be
  inaccurate and may lead to an erroneous determination of blazar
  evolution. This effect is more severe for the BL Lac class, since
  FSRQs can be misclassified as BL Lacs especially at high redshifts
  ($z>0.7$), where the latter are extremely rare.
\end{abstract}


\section{Introduction}

Blazars are active galactic nuclei (AGN) with strong jets forming a
small angle w.r.t. the line of sight and emitting variable, non-thermal
radiation across the entire electromagnetic spectrum
\citep{BR78,UP95}. Although intrinsically rare, blazars are being
detected in increasingly larger numbers by extragalactic surveys. In
fact, although only $3,500$ blazars are currently known
\citep{Massaro+09,Massaro+15}, their number is steadily growing thanks to, e.g.,
the {\it Fermi} \citep{Abdo10a, Abdo10b, Abdo11}, the optical Sloan
Digital Sky Survey \citep[SDSS:][]{Plotkin+10}, and the Planck
\citep{Pla11} surveys. Some faint blazars are also being detected as
serendipitous sources in Swift-XRT images \citep{Tu10}.

Blazars come in two main subclasses, whose major difference lies in
their optical properties: 1) Flat Spectrum Radio Quasars (FSRQs),
which show strong, broad emission lines in their optical spectrum,
just like radio quiet QSOs; and 2) BL Lacs, which are instead
characterized by an optical spectrum, which at most shows weak
emission lines, sometimes displays absorption features, and in many
cases can be completely featureless. Historically, the separation
between BL Lacs and FSRQs has been made at the (rather arbitrary)
equivalent width (EW) value of 5 \AA, rest-frame by \cite{Stickel+91}
and observed-frame by \cite{Stocke+91}. However, the search for a
possible bimodal distribution in the EW of the broad lines of radio
quasars has been unsuccessful. Indeed, \cite{SF97} pointed out that
radio-selected BL Lacs were, from the point of view of the emission
line properties, very similar to FSRQs but with a stronger
continuum. Most BL Lacs selected in the X-ray band, on the other hand,
had very weak, if any, emission lines, and \cite{Stocke+91}, when
studying the properties of the X-ray selected Einstein Medium
Sensitivity Survey (EMSS) sample, had to introduce another criterion
to identify BL Lacs, this time to separate them from normal
galaxies. This was based on the Ca H\&K break, a stellar absorption
feature typically found in the spectra of elliptical galaxies at
$\lambda \sim 4,000$ \AA. Given that its value in non-active
ellipticals is $\sim 50\%$, \cite{Stocke+91} chose a maximum value
of 25\% to ensure the presence of a substantial non-thermal continuum
superposed to the host galaxy spectrum. This was later revised to 40\%
\citep{Marcha+96,Landt+02}.

Based on all the above, it is clear that blazar classification depends
then on the details of their appearance in the optical band where they
emit a mix of three types of radiation: 1) a non-thermal, jet-related,
component; 2) thermal radiation coming from the accretion onto the
supermassive black hole and from the broad line region (at least in
most radio selected sources); 3) light from the host (giant
elliptical) galaxy. The strong non-thermal radiation, the only one
that spans the entire electromagnetic spectrum, is composed of two
broad humps, a low-energy one attributed to synchrotron radiation, and
a high-energy one, usually thought to be due to inverse Compton
radiation \citep{Abdo10c}.

Motivated by this observational background, \citeauthor{Gio+12} (2012, see
also \citealt{Gio+13}; \citealt{PG15}) run simulations
to try to explain the many selection effects that bias the FSRQ/BL Lac
classification. These simulations included, amongst other ingredients,
the blazar luminosity function and evolution, the thermal, non-thermal
and host galaxy components, and the peak frequency of the synchrotron
emission and EW distributions. Once the simulated samples had
been produced, the bias effects were introduced, in order to compare
the simulations with the real samples. These biases are basically the
limiting fluxes, to assess whether or not a source is detected by a
specific survey, and the {\it diluted} EW, which depends (apart from
the {\it intrinsic} EW) on the shape and intensity of the simulated
non-thermal emission and the host galaxy light, and determines whether
a source gets classified as a BL Lac or an FSRQ.  One of the results of this analysis
was that what we call a BL Lac can either be an object with a strong
non-thermal continuum (possibly due to strong Doppler factors), which
heavily dilutes its lines below the (arbitrary) dividing value of
EW $=5$~\AA, or a source with intrinsically weak emission lines. Radio
surveys tend to preferentially select the first type of sources (as
these are intrinsically radio powerful), while X-ray surveys select
mainly sources with lower radio power, related to low-ionization radio
galaxies (LERGs) and therefore optically featureless. The consequence
of this scenario is that the first kind of sources are classified as
BL Lacs only because they have a strong non-thermal emission, or a
high Doppler boosting, which swamp their emission lines. Such objects
are intrinsically FSRQs and so they should be classified. On the other
hand, most X-ray selected BL Lacs are truly (or nearly) featureless
and constitute the genuine BL Lac class \citep{Gio+12}.

The aim of this work is to test these predictions through X-shooter
optical and infrared spectroscopy of a sample of BL Lacs. The paper is
organized as follows: in Sect. 2 we describe the criteria adopted to
build our sample; in Sect. 3 we deal with the X-shooter observations
and the data reduction process; in Sect. 4 we present our results;
finally in Sect. 5 we discuss our findings and draw our conclusions.

\section{Sample selection}

An effective way of searching for FSRQs that have been erroneously
misclassified is to observe BL Lacs at $z > 0.7$ in the infrared
band. This is because at these redshifts one of the most prominent
emission lines in blazars, H$\alpha$ (rest-frame wavelength
$\lambda_{H\alpha} = 6562.8$\AA) moves out of the wavelength range
where optical spectroscopy is normally done.  In this scenario, some
blazars might have been classified as BL Lacs only because one of
their strongest and less contaminated emission lines falls beyond the
spectroscopic range normally used to classify blazars.

To verify this possibility, we selected a small sample of BL Lacs to
be observed at infrared wavelengths as follows: we started from the
BZCAT catalogue \citep{Massaro+09} and singled out all radio-selected
BL Lacs with good visibility from Paranal ($\delta <25^{\circ}$), $z >
0.7$, and a firm redshift identification (at least two lines detected
in emission, \citealt{SFK93,RS01,STF05}). We then removed
the objects for which the H$\alpha$ line was expected to fall in
regions dominated by atmospheric absorption features. Since the observed
redshift distribution of BL Lacs peaks at very low redshifts, our
final sample includes only five objects, which are listed in Table
1. Three of these have a redshift $z \sim 0.7-1$, and the H$\alpha$
line falls in the J band, while the other two have a redshift $z \sim
1.2 - 1.3$ and the H$\alpha$ is in the H band.

\section{X-shooter observations and data reduction}

We observed our blazar sample with X-shooter \citep{Dodo+06,
  Vernet+11}, a single-object medium resolution
($R=\lambda/\Delta\lambda = 4000\mbox{--}14000$) echelle spectrograph
mounted at the VLT-UT2 telescope. The observations were carried out
under programme 091.B-0092.  The choice of X-shooter as operating
instrument stems from its capabilities to perform simultaneous optical
and IR spectroscopy with a single exposure. This way we can also study
the ratio between optical and IR band lines without being affected by
the variability typical of blazars.

For both observations the slit width was set to 0.9\arcsec{} in the
visual (VIS) and near-infrared (NIR) arms, and 1.0\arcsec{} in the
ultra-violet and blue (UVB) arm. The UVB and VIS CCD detectors were
rebinned to $1 \times 2$ pixels (binned in the spectral direction but
not in the spatial one) to reduce the readout noise.  With this
configuration, the nominal resolution is different for the three arms:
$R \sim 5100$, $8800$, $5300$ for the UVB, VIS, and NIR arms,
respectively.

All our targets were observed during ESO period P91, i.e., between 1
April and 30 September 2013. The list of observations is reported in
Table 1, together with the observing dates and the total exposure
times. Sources requiring more than 1 hour of integration were observed
more than once. Each observation is constituted by several exposures,
taken nodding along the slit with an offset of $5\arcsec$ between
exposures, following a standard ABBA pattern.

We processed the spectra using version 1.4.5 of the X-shooter data
reduction pipeline \citep{Goldoni+06,Modigliani+10}.  The
pipeline carries out the following steps: the raw frames are first
bias-subtracted, and cosmic ray hits are detected and removed using
the method developed by \cite{vanDokkum01}. The frames are divided by a
master flat field obtained using day-time flat field exposures with
halogen lamps. The orders are extracted and rectified in wavelength
space using a wavelength solution obtained from day-time calibration
frames. The resulting rectified orders are shifted and co-added to
obtain the final two-dimensional spectrum. In the overlapping regions,
orders are merged by weighing them using the errors propagated during
the reduction process. From the resulting two-dimensional merged
spectrum, a one-dimensional spectrum with the corresponding error file
and bad pixel map was extracted at the source position. Telluric
corrections were applied for spectra whose features fall in regions
strongly contaminated by sky lines. The
IRAF\footnote{http://iraf.noao.edu/} task {\it telluric} was adopted
to perform this correction.  Flux calibration is not required by the
goal of our work, since we are interested only in line
spectroscopy. Normalized spectra have thus been produced and used
throughout this work.

\begin{table*}
\caption{\bf X-shooter observations}
\centering
{\footnotesize
\smallskip
\begin{tabular}{|l|c|c|c|c|c|c|c|}
  \hline 
  Source & RA (J2000, hr)      & Dec (J2000, deg)        & redshift&    Observation date         & Exposure time (s)& S/N  & airmass     \\
  \hline                                             
  BZBJ0141$-$0928 & 01 41 25.83 & -09 28 42.9 &$0.733^{(1)}$       &    2013-07-24T08:39:08.945  &2880 &  $\sim 20$  & $1.09$ \\
                &            &             &                   &    2013-08-17T04:57:54.150  &2880 &  $\sim 20$  & $1.56$ \\
                &            &             &                   &    2013-08-19T04:55:01.183  &2880 &  $\sim 20$  & $1.52$ \\
                &            &             &                   &    Total                    &8640 &  $\sim  35$ &        \\
  \hline     
  BZBJ0238+1636 & 02 38 38.93 & +16 36 59.0 &  $0.94^{(3)}$      &    2013-08-10T09:25:40.669  &1800 & $\sim 12$  &$1.34$\\
  \hline
  BZBJ2031+1219 & 20 31 54.99 & +12 19 41.0 & $1.215^{(2)}$      &    2013-08-10T02:38:58.540  &2880 & $\sim 8$    & $1.30$ \\
                &            &             &                   &    2013-08-10T03:48:49.421  &2880 & $\sim 8$    & $1.26$ \\
                &            &             &                   &    Total                    &5760 & $\sim 12$   &  \\  
  \hline 
 BZBJ2134$-$0153  & 21 34 10.30 & -01 53 17.0 & $1.283^{(1)}$      &    2013-05-27T06:29:07.736  &3600 & $\sim 15$  &$1.54$\\ 
  \hline
  BZBJ2243$-$2544 & 22 43 26.40 & -25 44 30.0 & $0.774^{(4)}$      &    2013-06-20T08:33:30.168  &1800 & $\sim 35$  &$1.02$\\
  \hline                                             
\end{tabular}
}

$^{(1)}$ \cite{RS01}; $^{(2)}$ \cite{SK93}; $^{(3)}$ \cite{Healey+08}; $^{(4)}$ \cite{SFK93}
\end{table*}

\begin{table*}
\caption{\bf Rest-frame equivalent widths for our blazar sample.}
\centering
{\footnotesize
\smallskip
\begin{tabular}{|l|c|c|c|c|c|}
\hline 
                              &  BZBJ0141$-$0928     & BZBJ0238+1636       & BZBJ2031+1219           & BZBJ2134$-$0153       & BZBJ2243$-$2544     \\
\hline                                                                                           
Line                          & rest-frame EW (\AA)  & rest-frame EW (\AA) &rest-frame EW (\AA)      &  rest-frame EW (\AA)  &rest-frame EW (\AA)  \\
\hline                                                                                           
{Mg}{II}$\lambda$2800\AA & $1.0\pm0.2$          &$1.5 \pm 0.4$       & Tentative               & $< 0.3$               & $< 0.2$ \\              
\hline                                                                                           
[{O}{II}]$\lambda$3728\AA & $0.3\pm0.2$          &$0.5 \pm 0.4$        &$<0.2$                   & $2.2\pm0.3$           & $0.6\pm0.2$ \\              
\hline                                                                                           
H$\beta$                      & $<0.2$               &$ < 1.0 $            & $<0.3$                  & $<0.4$                & $<0.2$ \\              
\hline                                                                                           
[{O}{III}]$\lambda$4959\AA& $0.3\pm0.1$          &$0.6\pm0.3$          & $<0.3$                  & $<0.3$                & $0.2\pm 0.2$ \\              
\hline                                                                                           
[{O}{III}]$\lambda$5007\AA& $1.0\pm0.2$          & $1.5\pm 0.4$        & $<0.3$                  & $2.8\pm0.6$           & $0.5\pm0.2$ \\              
\hline                                                                                           
H$\alpha$                     & $<1.2$               &$6.7\pm0.4$          & $<0.5$                  & $3.2\pm0.4$           & $2.5\pm0.5$ \\              
\hline  
\end{tabular}
}
\end{table*}

\section{Results}

We computed the rest-frame EWs for the emission lines of our sources,
which are given in Table 2. The EWs have been computed through direct
line integration. Errors have been estimated resampling the spectra
and take into account both the normalization uncertainties and the S/N
of the spectra. We discuss the sources individually in the following
sub-sections.

\subsection{BZBJ0141$-$0928 (PKS 0138$-$097)}

\begin{figure}
\centering
\includegraphics[angle=-0,width=9cm]{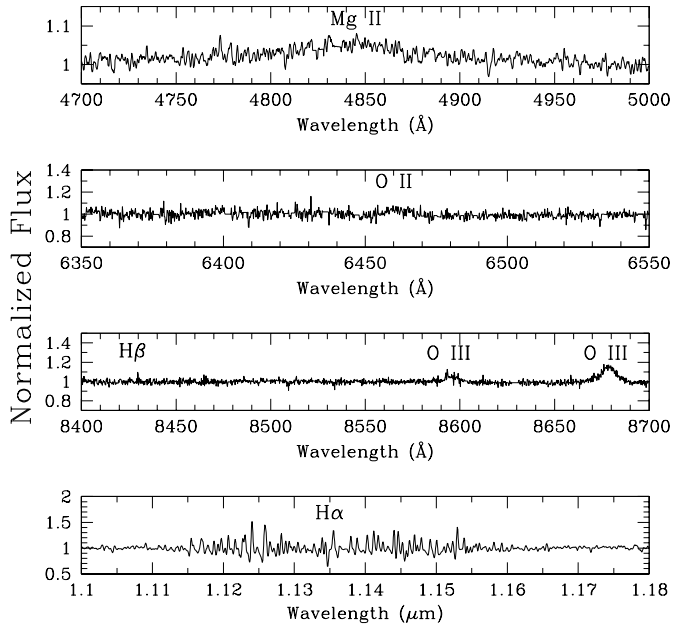}
\caption{The emission lines of BZBJ0141$-$0928.}
\label{fig1}
\end{figure}

For this blazar we detected the {Mg}{II} doublet at $2800$\AA\,
and three oxygen lines, namely, [{O}{II}] and the [{O}{III}]
doublet (see Fig. \ref{fig1}). A redshift of $z=0.736\pm0.001$ is
derived, in agreement with that reported by \cite{RS01}.
Despite a very high S/N ratio, we can not detect Hydrogen emission
lines, but the H$\alpha$ feature falls in a region contaminated by
several sky lines.

An intervening system is detected at $z=0.5005 \pm 0.0005$, featuring
absorption by the {Mg}{II} $\lambda\lambda$2796,2803,
{Fe}{II}$\lambda\lambda$2586,2600 doublets, and the
{Fe}{II}$\lambda$2382 line.

\subsection{BZBJ0238+1636 (AO 0235+164)}

\begin{figure}
\centering
\includegraphics[angle=-0,width=9cm]{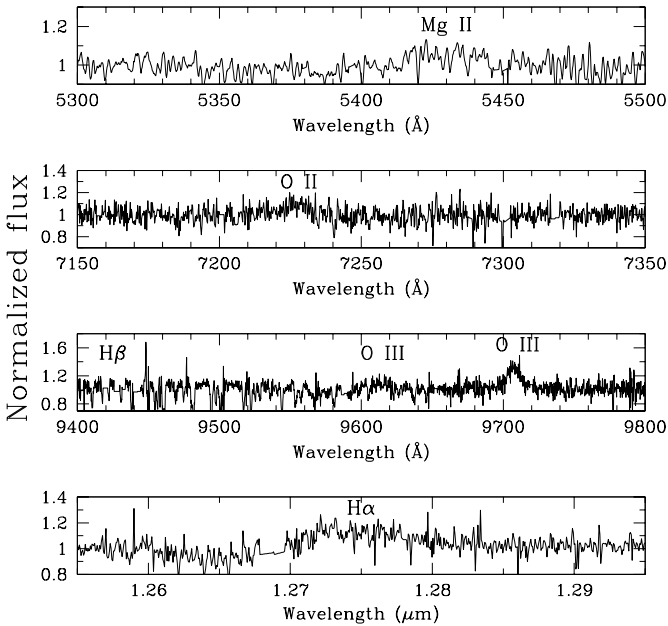}
\caption{The emission lines of BZBJ0238+1636.}
\label{fig2}
\end{figure}

This is the blazar featuring the most prominent emission lines in our
sample. We confirm the redshift of this source to be $0.942 \pm 0.001$
\citep{Healey+08}. Five lines have been detected, namely, the
{Mg}{II} doublet, [{O}{II}], the [{O}{III}] doublet and
H$\alpha$. The H$\beta$ line falls in a telluric region, so only a
shallow upper limit could be set. The H$\alpha$ line rest-frame EW is
higher than $5$ \AA. All the other features detected in the optical
band have instead values $<5$\AA. BZBJ0238+1636, previously classified
as a BL Lac due to the EW of its optical lines, should be
re-classified as an FSRQ according to its strong H$\alpha$ feature in
the IR band. The detected emission lines are displayed in
Fig. \ref{fig2}.

\cite{Raiteri+07} have carried out a spectroscopic monitoring of
this source, finding that the EW of the {Mg}{II} line was strongly
dependent on its magnitude, being smaller when the source was
brighter. Our rest-frame {Mg}{II} EW $\sim 1.5$ \AA~is consistent
with this trend, as this object was in a relatively bright state at
the time of our observations ($R \sim
17$\footnote{http://www.bu.edu/blazars/VLBAproject.html}).

\subsection{BZBJ2031+1219 (PKS 2029+121)}

This blazar is almost completely featureless. We only report a
tentative detection of the {Mg}{II} doublet at the redshift $z=1.215$,
consistent with that by \cite{SK93}. In addition, we detect the
intervening absorber seen by \cite{Bergeron+11} at $z=1.116$, which
features the {Mg}{II}$\lambda,\lambda2796,2803$ doublet and the
[{Mg}{I}$\lambda2852$] line, which gives more robustness to our
redshift determination. Since the main emission lines should fall in
regions relatively free of sky lines, this might suggest that we have
caught BZBJ2031+1218 in a high state, during which the thermal
emission is overwhelmed by the non-thermal one. If this is the case,
the emission lines are diluted by the synchrotron emission below the
detection threshold of our spectrum. Only upper limits for [{O}{II}],
the [{O}{III}] doublet, H$\beta$ and H$\alpha$ have been set at the
redshift of $z=1.215$ (see Table 2).

\subsection{BZBJ2134$-$0153 (4C$-$02.81)}

\begin{figure}
\centering
\includegraphics[angle=-0,width=9cm]{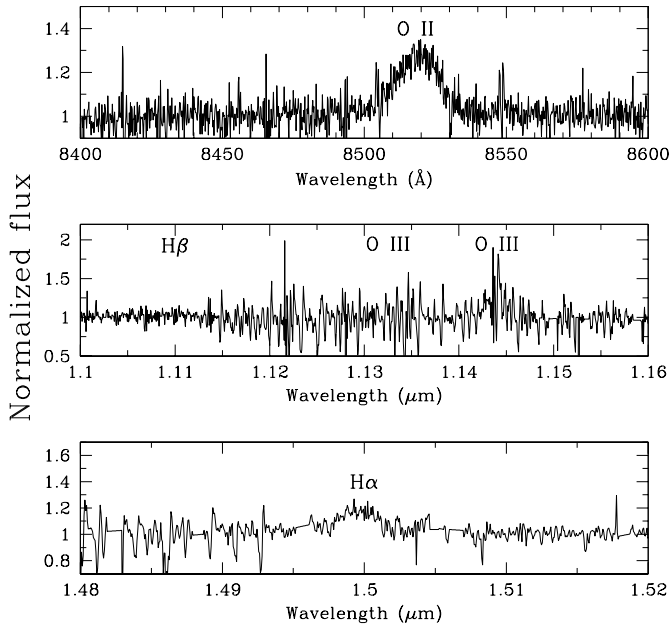}
\caption{The emission lines of BZBJ2134$-$0153.}
\label{fig3}
\end{figure}

This blazar shows three emission lines, namely, [{O}{II}],
[{O}{III}]$\lambda$5007\AA~and H$\alpha$. We found a redshift of
$z=1.286 \pm 0.001$, consistent with that reported by \cite{RS01}. For
the H$\beta$ and [{O}{III}]$\lambda$4959\AA~ features we could set
only upper limits. However, the [{O}{III}] doublet falls in a region
highly contaminated by sky lines. It is interesting to note that the
H$\beta$ lower limit is about eight times smaller than the H$\alpha$
value, despite the ratio between the two oscillator strengths being
only $\sim 5$. We do not detect the {Mg}{II} emission line reported in
\cite{Bergeron+11} nor the {Mg}{II} intervening absorber at
$z=1.2458$. However, the reported EW of the intervening {Mg}{II}
system ($\sim 0.2$\AA) is below our detection threshold. The detected
emission lines are displayed in Fig. \ref{fig3}.

\subsection{BZBJ2243$-$2544 (PKS 2240$-$260)}

\begin{figure}
\centering
\includegraphics[angle=-0,width=9cm]{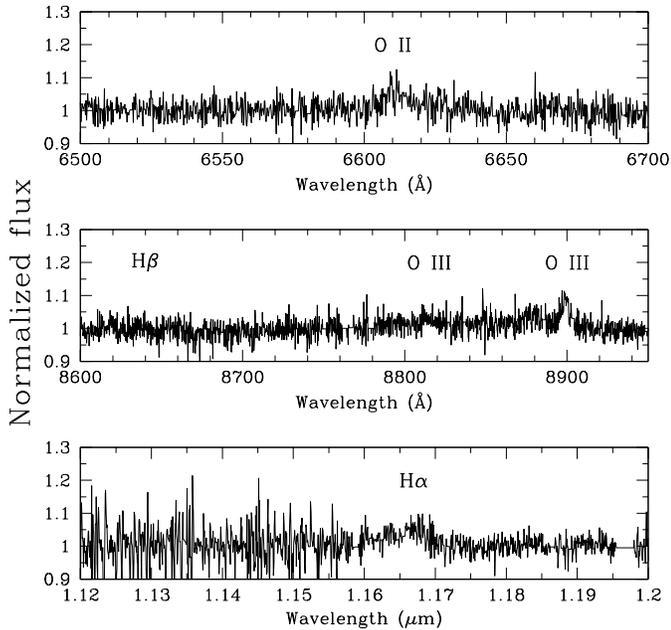}
\caption{The emission lines of BZBJ2243$-$2544.}
\label{fig4}
\end{figure}

This blazar shows four emission lines, namely [{O}{II}], the
[{O}{III}] doublet and H$\alpha$. The redshift determined for this
object is $z=0.780\pm0.002$, consistent with \cite{SFK93}. For the
H$\beta$ feature only an upper limit could be set. Again, we note that
this lower limit is more than ten times lower than the H$\alpha$
value, despite the ratio between the two oscillator strengths is just
a factor of $\sim 5$. The detected emission lines are displayed in
Fig. \ref{fig4}.

 \section{Discussion and conclusions}

 We have looked for H$\alpha$ emission in the IR spectra of five BL
 Lacs at $z > 0.7$. We have found that two sources (BZBJ0238+1636,
 rest-frame EW $> 5$~\AA~and BZBJ2134$-$0153, observed EW $> 5$~\AA)
 could be classified as FSRQs by one of the classification criteria
 that have been used in the literature. Indeed, BZBJ0238+1636 was
   defined as an ``intruder'' BL Lac in \cite{Ghisellini+11}. This is
   because its high $\gamma$-ray luminosity and Eddington ratio
   resemble those of typical FSRQs, despite its $\gamma$-ray energy
   index being border-line between the two classes. Concerning
   BZBJ2134$-$0153, its classification is ambiguous, since it was
   considered an FSRQ in the 1LAC catalog \citep{Abdo10d} and a BL Lac
   in the 2FGL one \citep{Nolan+12}. Its $\gamma$-ray luminosity of
   $\sim 10^{47}$ erg s$^{-1}$ and its WISE colors (see
   \citealt{Massaro+11}) better place this object in the FSRQ
   class. In addition, given the steep NIR-to-UV slope, if the
   redshift of this object were $z\sim2$, a strong Ly-$\alpha$ line
   would have been detected, definitely placing BZBJ2134$-$0153 in the
   FSRQ class. A third source in our sample shows border-line
 (observed) EW (BZBJ2243$-$2544, EW = $4.4\pm0.5$~\AA). The remaining
 two objects show very weak lines or nearly featureless spectra.

 Our sample shows that, whenever emission lines are present, H$\alpha$
 is always the most prominent one (the only exception being
 BZBJ0141$-$0928, for which it falls in a region strongly contaminated
 by sky lines). This appears to be the case even if this feature falls
 relatively close to the peak of the synchrotron emission in our
 sources, which are low synchrotron-peaked BL Lacs.  In this situation
 the synchrotron emission, which is non-thermal, should strongly
 dilute the thermal processes such as the H$\alpha$ emission
 lines. H$\alpha$ turns then out to be very important if one has to
 rely on a classification scheme that discriminates BL Lacs from FSRQs
 according to the EW of their lines. In this context IR spectroscopy
 turns out to be vital for relatively high redshift ($z > 0.7$)
 sources.
 
 The danger of source misclassification based solely on EW strength
 has been recently stressed also by \cite{Ruan+14}. They studied
 multi-epoch spectra of more than 300 blazars, finding that for six of
 them the EW of their emission lines crossed the $5$ \AA\, dividing
 line in different observations. These objects have high accretion
 rates, strong variability both in the thermal and non-thermal
 continuum and a synchrotron peak frequency similar to that of
 FSRQs. \cite{Ruan+14} propose that these sources are likely FSRQs
 whose jet axis points extremely close to the line of sight. This
 peculiar geometry causes the emission lines to be strongly diluted by
 the non-thermal continuum when the latter is particularly
 strong. They conclude, similarly to \cite{Gio+12}, that a simple
 classification based on the EW strength may lead to consider these
 FSRQs as BL Lacs and to a wrong evaluation of the BL Lac evolution,
 especially at high redshifts. This effect could be even more severe
 when considering that blazars with redshift larger than $0.7$,
 currently classified as BL Lacs, could be in fact FSRQs with the
 H$\alpha$ emission line in the IR window. Thus, a new, EW-independent
 operational definition for "BL Lac object" is needed.

 Such a picture has also strong implications on blazar classification
 and evolution and on the so-called ``blazar sequence'', and predicts
 the association of ``real'' BL Lacs with LERGs and ``fake'' BL Lacs
 with high-ionization radio galaxies (HERGs). In particular, the
 evolution derived for FSRQ samples in the radio, and also
 $\gamma$-ray, band could be biased by the exclusion of relatively
 high-redshift misclassified sources. Given the role that radio
 emission is thought to play in galaxy evolution through the so-called
 ``radio mode'' accretion \citep{Croton+06}, this is also relevant
 for ``AGN feedback'' via the selection of relatively high-redshift
 HERGs but especially LERGs, which being intrinsically less powerful
 are harder to identify, already at moderately high redshifts, through
 blazar samples.

\section*{Acknowledgments}
We thank the referee and the editors for helpful comments. This work
is based on observations made with ESO telescopes at the La Silla
Paranal Observatory under programme ID 091.B-0092.

\label{lastpage}
\end{document}